\documentclass[]{aa}
\usepackage{graphicx}
\usepackage{subfigure}

\begin{document}
%\thesaurus{}

\title{Far Infrared mapping of NGC~891\thanks{Based on
   observations with the {\it Infrared Space Observatory (ISO)}, an ESA project
   with instruments funded by ESA member states (especially the PI countries:
   France, Germany, the Netherlands, and the United Kingdom) and with
   the participation of ISAS and NASA.}}
\author{Cristina C. Popescu\inst{1,2}, Richard. J. Tuffs\inst{1}, 
Nikolaos D. Kylafis\inst{3,4}, Barry F. Madore\inst{5,6}}

\offprints{Cristina.Popescu@mpi-hd.mpg.de}
\institute{Max-Planck-Institut f\"ur Kernphysik,
           Saupfercheckweg 1, D-69117 Heidelberg\\
           \email{Cristina.Popescu@mpi-hd.mpg.de}\\
           \email{Richard.Tuffs@mpi-hd.mpg.de} 
           \and
           Research Associate, The Astronomical Institute of the Romanian
           Academy, Str. Cu\c titul de Argint 5, Bucharest, Romania
           \and University of Crete, Physics Department, P.O. Box 2208, 710 03
           Heraklion, Crete, Greece
            \and Foundation for Research and Technology-Hellas, 
             71110 Heraklion, Crete, Greece
           \and NASA/IPAC Extragalactic Database, 770 S. Wilson Avenue, 
           Pasadena, California 91125, USA
           \and The Observatories of the Carnegie Institution of Washington, 
           813 Santa Barbara Str., Pasadena, 91101 California, USA}
\date{Received; accepted}

\abstract{We present deep maps of dust emission from the edge-on spiral 
galaxy NGC~891, obtained with the ISOPHOT instrument on board the 
Infrared Space Observatory in broad band filters with reference wavelengths 
centered on  170 and 200\,${\mu}$m. Using new processing methods to remove the 
effects of detector transients from the data, we detect cold dust at high 
dynamic range. 
The observed surface brightness distribution and colour profile of the 
far-infrared (FIR) emission are found to be in good agreement with predictions
for their counterparts derived from the model of Popescu et al. (2000a). 
Thus, NGC~891 is the first galaxy for which an intrinsic distribution of dust 
and stars could be found which simultaneously accounts for both the
optical/near-IR  and FIR morphologies.
\keywords{galaxies: individual: NGC~891, galaxies: spiral, 
galaxies: structure, ISM: dust, infrared: continuum, radiative transfer}
}
\authorrunning{Popescu et al. 2003}
\maketitle

\section{Introduction}

NGC~891 is one of the most extensively observed and studied edge-on spiral
galaxy in the nearby universe. At a distance of 9.5\,Mpc 
(van der Kruit \& Searle 1981), it has been classified as an Sb galaxy by 
Sandage (1961) and is often quoted to be very similar to our own Galaxy. 
However it has also some unusual characteristics. For example NGC~891 contains
one of the most spectacular layers of extraplanar diffuse ionised gas (DIG) 
(Dettmar 1990; Rand et al. 1990; Keppel et al. 1991; Pildis et al. 1994; 
Rand 1997, 1998; Hoopes et al. 1999; Howk \& Savage 2000, Otte et al. 2001).
The galaxy also has a radio continuum halo (Allen et al. 1978; Hummel et al. 
1991), an HI halo (Swaters et al. 1997) and an X-ray halo 
(Bregman \& Pildis 1994; Bregman \& Houck 1997). 

The disk of NGC~891 is a strong source of both CO and $^{13}$CO emission
(Sofue et al. 1987, Garc\'{\i}a-Burillo et al. 1992) and the global
distribution and kinematics of molecular gas have been investigated by 
Handa et al. (1992),
Scoville et al. (1993), Garc\'{\i}a-Burillo \& Gu\'elin (1995), Sakamoto et 
al. (1997). The first extragalactic direct detection of large-scale molecular
hydrogen was established in the disk of NGC~891 (Valentijn \& van der Werf
1999), based on observations with the Short-Wavelength Spectrometer (SWS)
instrument aboard the Infrared Space Observatory (ISO).  The spectrum of the
unidentified infrared (UIR) emission bands between 5.9 and 11.7\,${\mu}$m has
been also observed for the first time in the disk of an external galaxy in
NGC~891 (Mattila et al. 1999), using the low-resolution
spectrometer of the ISOPHOT instrument aboard ISO. The mapping of the
Unidentified Infrared Bands emitted by
NGC~891 was also done with the ISOCAM instrument  on board ISO (Le Coupanec et
al. 1999).

\begin{table*}[htb]
\caption{Log-book of the observations}
\begin{tabular}{lccccccccc}
\hline\hline
Field & Filter & TDT$^{1}$ & \multicolumn{2}{c}{Map Centre (J2000)} & PA$^{2}$
& \multicolumn{2}{c}{Map Sampling} & \multicolumn{2}{c}{Background}\\  
      &        &         & RA          &   DEC      &         
& \multicolumn{2}{c}{Y x Z$^{3}$}       & ISO    & COBE\\
      &        &         &             &            &  deg    
& arcsec                           & arcsec & MJy/sr & MJy/sr\\ 
\hline
Centre & C160 & 65600207 & 02 22 33.03 & +42 20 55.5 & 148.82 & 30.66 & 46.00 
& -     &  -\\
South  & C160 & 61100404 & 02 22 11.47 & +42 12 04.3 & 168.68 & 30.66 & 31.00 
& 11.25 & 9.17\\
North1 & C160 & 65600403 & 02 22 54.79 & +42 29 48.5 & 148.79 & 30.66 & 91.97 & 11.65 & 8.68\\
North2 & C160 & 80802313 & 02 22 54.79 & +42 29 48.5 & 341.63 & 30.66 & 91.99 
& 12.75 & 10.26\\
Centre & C200 & 65600207 & 02 22 33.03 & +42 20 55.5 & 148.83 & 30.66 & 46.00 
&  -    &  -\\
South  & C200 & 61100404 & 02 22 11.47 & +42 12 04.4 & 168.73 & 30.67 & 31.00 
&  4.87 & 8.13\\
North1 & C200 & 65600403 & 02 22 54.79 & +42 29 48.5 & 148.79 & 30.67 & 91.99 
&  6.54 & 7.72\\
North2 & C200 & 80802313 & 02 22 54.79 & +42 29 48.5 & 341.64 & 30.66 & 92.00 
&  7.45 & 9.01\\
\hline
\end{tabular}

$^{1}$ Target Dedicated Time identifier. The first three digits give the
orbit identifier, which is also the epoch of the observation in days after 
November 17th 1995.\\
$^{2}$ positive Y direction (the direction of the chopper sweep), degrees E 
from N\\
$^{3}$ spacecraft coordinates\\
\end{table*}

The distribution of the cold dust in NGC~891 has been observed at
submillimeter(submm)/millimeter(mm) wavelengths using the IRAM 30\,m 
telescope 
(Gu\'elin et al. 1993) and the Submillimeter Common-User Bolometer Array 
(SCUBA) at the James Clerk Maxwell Telescope (JCMT)(Alton et al. 1998; 
Israel et al. 1999). The deep SCUBA maps revealed dust 
emission over 2/3 of the optical disk, but did not shed light on the 
existence of an 
extraplanar dust emission (Alton et al. 2000). Large amount of cold dust 
($\sim 15$\,K) was found in the disk (Alton et al. 1998) - an order of 
magnitude more than the amount of warm dust detected by 
Wainscoat et al. (1987) using the Infrared
Astronomical Satellite (IRAS).

However, the peak of the spectral energy distribution (SED) from 
cold dust, lying at the long far-infrared (FIR) wavelengths, is outside the 
wavelength coverage of IRAS, and thus has not been readily accessible until
recently. With a wavelength coverage extending to 240\,${\mu}$m and a superior 
intrinsic sensitivity as compared to IRAS, the ISOPHOT instrument 
(Lemke et al. 1996) on board ISO (Kessler et al. 1996) was the first to 
directly measure the peak of the FIR SED for a number of nearby galaxies 
(Tuffs et
al. 1996 for NGC~6946, Haas et al. 1998 for M~31, Wilke et al. 2003 for the 
SMC, Hippelein et al. 2003 for M~33), 
for smaller statistical samples (Kr\"ugel et al. 1998; 
Siebenmorgen et al. 1999; 
Contursi et al. 2001), for the ISOPHOT Virgo Cluster Deep Sample 
(Tuffs et al. 2002a,b; Popescu et al. 2002; Popescu \& Tuffs 2002a) and for 
the sample of Bright Revised Shapley Ames galaxies (Bendo et al. 2002, 2003).

\begin{figure*}[htb]
\includegraphics[scale=0.55]{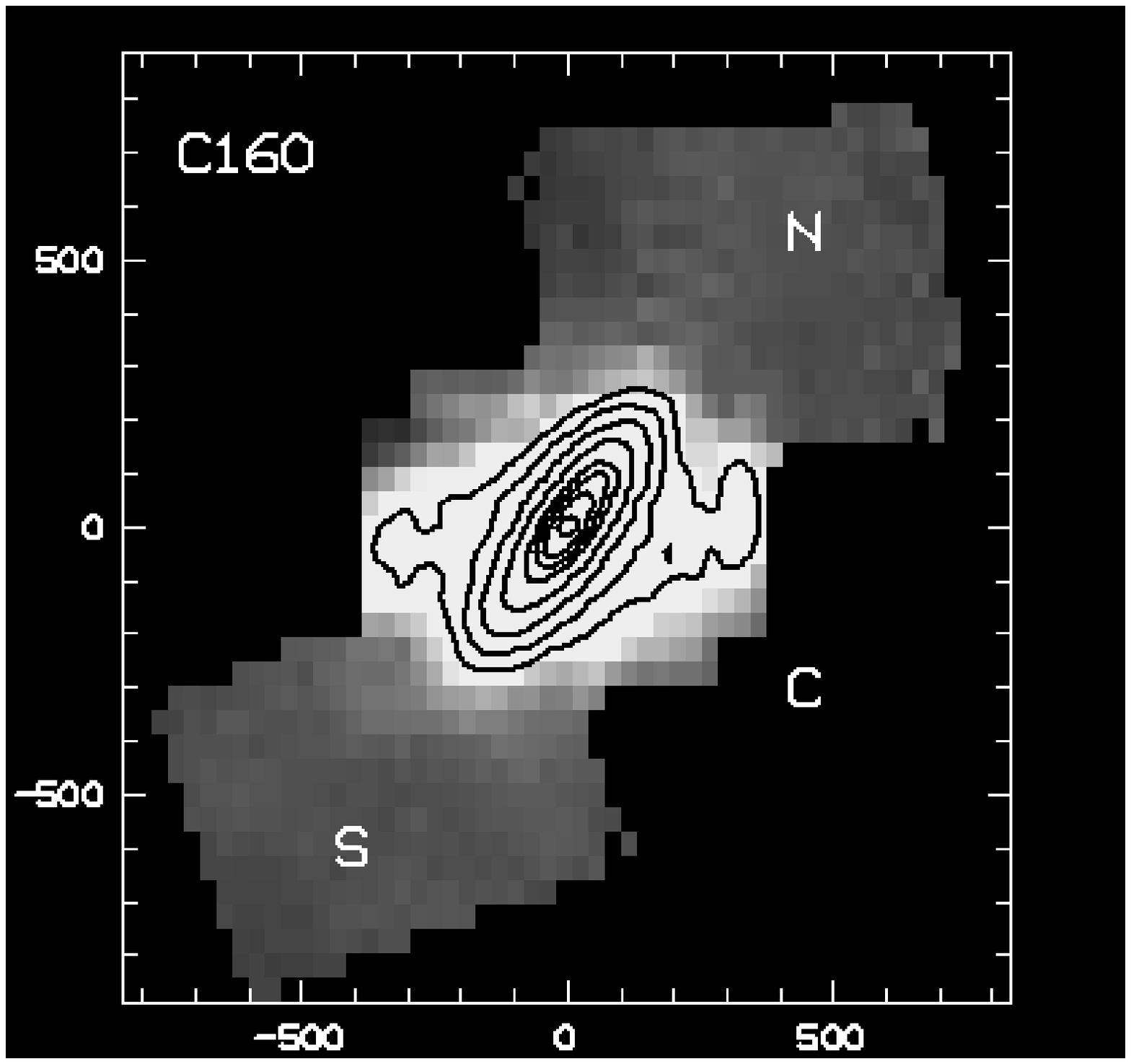}
\includegraphics[scale=0.55]{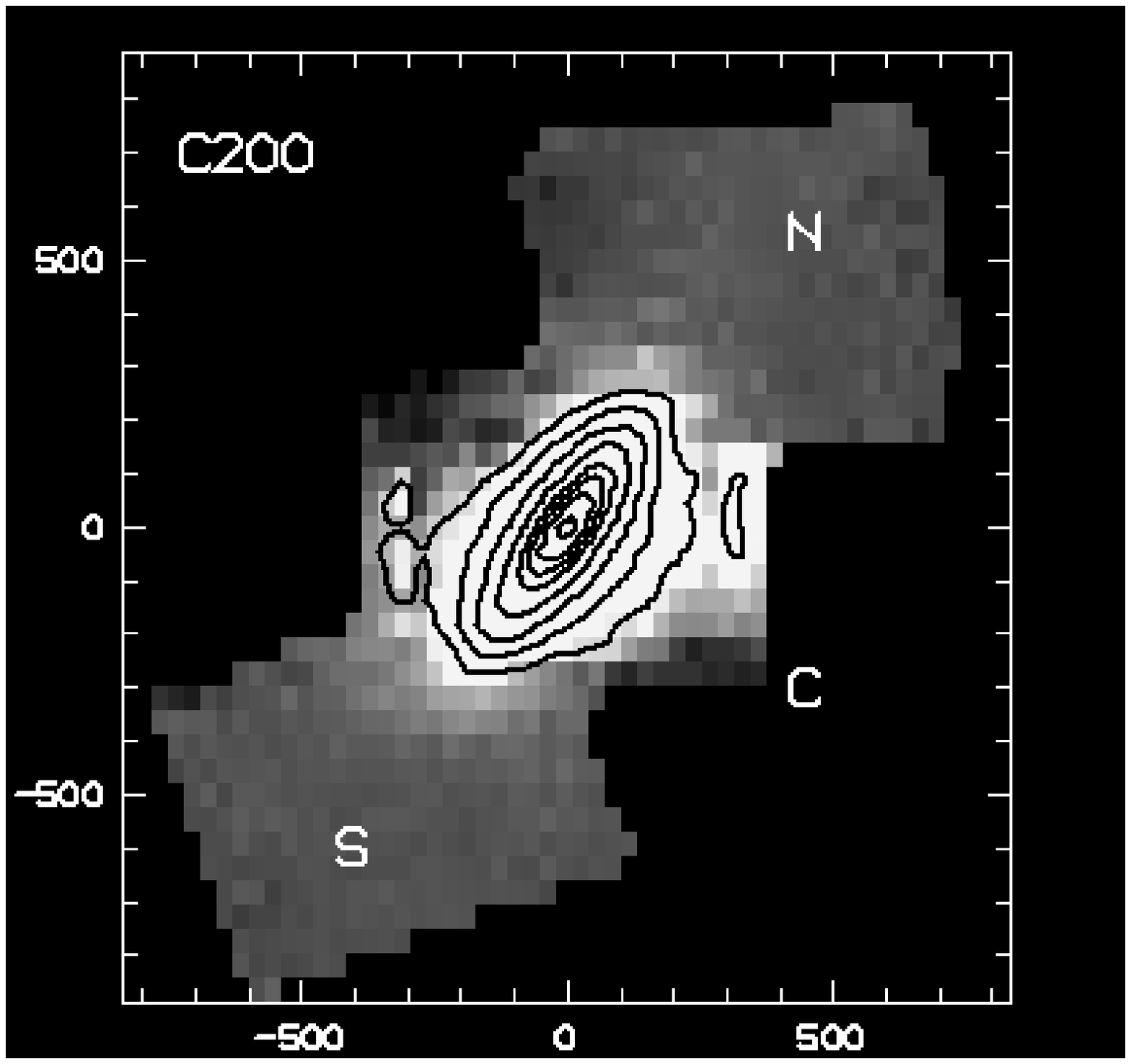}
\caption{Greyscale images of the C160 and C200 mosaic maps of NGC~891 with
coordinates given as offsets (in arcsec) from the centre of the galaxy. The 
orientation and sampling of the mosaic have been set to those
of the central fields given in Table~1. The 
southern, northern and central fields of the mosaic are marked with the 
letters S, N, and C, respectively. 
For the C160 image the contour levels are: 6.5, 13.4, 27.5, 56.5, 106.0, 139.2,
172.4 and 205.5 MJy/sr. For the C200 image the contour levels are: 6.5, 12.8,
25.3, 50.0, 87.3, 114.3, 141.2 and  168.1 MJy/sr. 
To show the full dynamic range of the images, down
to the noise level in the southern and northern fields,
the bright emission from the disk of NGC~891 is depicted as 
contours overlaid on a maximum (white) greyscale level set
to 2 percent of the peak brightness of the galaxy. The FWHM of the 
ISOPHOT beam is 91$^{\prime\prime}$ and 93$^{\prime\prime}$ at 170 and 
200\,${\mu}$m, respectively.}
\end{figure*}

Here we used the dedicated mapping mode P32 (Tuffs \& Gabriel 2003) of 
ISOPHOT to obtain deep maps of NGC~891 at 170 and 200\,${\mu}$m
wavelengths. These data, taken at the peak of the SED from cold dust, 
provide a precise measurement of the stellar light re-radiated by grains in 
NGC~891. They also allow cold dust to be probed in 
regions of low surface brightness, such as the disk periphery and halo of 
NGC~891, regions which are currently inaccessible to submm facilities.
When combined with the higher column density presented by the edge-on 
orientation, the ISOPHOT observations of NGC~891 have the capability of 
tracing cold dust to higher galactic radii. Thus, the goals of this study 
are: 1) to derive flux densities at longer FIR
wavelengths, and thereby to directly measure the peak of the FIR
SED in NGC~891; 2) to measure the physical extent of the dust disk and in 
particular to 
search deep for a cold dust counterpart to the extended HI disk; 3) 
to search for cold extraplanar FIR emission; 4) to compare the brightness and 
colour profile of the dust disk with predictions from the three dimensional 
model of stellar and dust distributions in NGC~891 proposed by 
Popescu et al. (2000a).
This paper concentrates on the 1st, 2nd, and 4th goals, while the search for a 
dust counterpart to the extended HI disk is presented in Popescu \& Tuffs 
(2003). In Sect.~2 we present the observations and
data reduction, including the derivation of the integrated flux densities
from the maps. In Sect.~3 we give a detailed comparison between the data
and the model predictions for NGC~891. Some implications of this 
comparison are discussed in Sect.~4. A summary is given in Sect.~5.

\section{Observations and data reduction}

The observations were made using the ISOPHOT-C200 2$\times2$ pixel array
in the C160 and C200 filters\footnote{We warn the reader that the name
of the C160 filter does not correspond to the central wavelength, unlike the
case of the C200 filter.}, which respectively cover passbands of
$130\,-\,218$ and $170\,-\,239\,\rm \mu m$ and have central
wavelengths of 170 and 200\,${\mu}$m. The ``P32'' mapping mode was 
used to provide near Nyquist sampling over a large area encompassing
the optical galaxy and the extended HI disk, as well as the surrounding
background. The HI disk extends $\sim10$ arcmin from the nucleus in the 
southern half of the galaxy and $\sim 7$ arcmin in the northern half 
(Swaters et al. 1997). In order
to cover the whole southern field, and also to have a symmetrical map, a field
of radius $\pm13.5$ arcmin ($\pm 40$\,kpc) was mapped along the major axis of 
the galaxy. 

A total of 19,000 seconds were spent in three overlapping 
fields: north, south and central. Due to scheduling constraints, the 
coverage of the northern field was made in two shorter overlapping 
observations at each wavelength. Thus, 4 maps were made at each 
wavelength. The parameters of each of these observations are summarised 
in Table~1. Part of these data sets have been presented by Popescu \& Tuffs
(2002b), Tuffs \& Popescu (2003) and by Dupac et al. (2003).

The data for each observation were separately processed using the 
latest P32 reduction package (Tuffs \& Gabriel 2003), which corrects 
for the transient response of the detector pixels. This allowed high 
dynamic range maps to be constructed to levels of 1 percent of the peak 
disk brightness. A time-dependent flat-field correction was made for 
each map, by fitting a cubic function to the response of the
detector pixels to the background. Calibration was made using V8.1 of
the PHOT Interactive Analysis (PIA) Package (Gabriel et al. 1997). We 
emphasise that although the maps are oversampled, independent data contribute 
to each map pixel.  

Finally, the maps were combined into a mosaic spanning half a degree 
parallel to the plane of the galaxy. To achieve a common photometric
scale over all the fields, the northern and southern fields were first 
scaled to the background brightness found from colour-corrected 
COBE/DIRBE maps within a 1.5 degree radius circle of NGC~891, taking 
into account the temporal variation of the zodiacal light component 
of the background. Values of the observed ISO backgrounds and the
COBE/DIRBE backgrounds to which they were scaled are given in Table~1.
Scaling factors between the central field and the southern and northern 
fields were then calculated from source structure in the 
overlapping regions of the maps, and applied to the data of the 
central field as a linear interpolation in position between the 
overlapping regions. Finally, backgrounds were subtracted from the 
southern and northern fields, and a constant background from the 
central field such that the absolute rms discrepancy in the overlap 
regions was minimised. 

Overall, the self consistency of the pixel 
responses to the backgrounds suggests a systematic uncertainty of 
5 and 10 percent in the C160 and C200 filters, respectively.
The random noise variations in the southern and northern
regions of the mosaic are 0.1 and 0.2 MJy/sterad, respectively,
some 1000 times fainter than the peak emission. Thus the maps
are dynamic-range limited by the sidelobes of the beam response
to the bright disk emission.

The derived mosaic maps at 170 and 200\,${\mu}$m are shown in Fig.~1. 
The central field is dominated by the FIR emission from the optical disk of the
galaxy. Fainter emission is seen out to at least the edge of the optical disk 
(360$^{\prime\prime}$ from the nucleus). A detailed investigation of the FIR
emission beyond the optical disk is presented in Popescu \& Tuffs (2003). 
Another feature seen in the maps is a pair of wings (highlighted by the
outer contours in Fig 1) separated from the 
nucleus by $\pm 320$ arcsec in the scan direction. The feature is more 
prominent at 170\,${\mu}$m than at 200\,${\mu}$m. The alignment of this 
feature with the scan direction points to an instrumental origin. However a 
specific cause could not be identified and it cannot be completely ruled out 
that extraplanar emission has not been detected.    
 
The integrated fluxes are given in Table~2, together with their uncertainties
($\epsilon(F)$), peak brightness ($B_{\rm peak}$) and the corresponding 
uncertainties (${\epsilon}(B_{\rm peak})$) for both C160 and C200 maps.
The last column represents an estimate of the systematic uncertainty in the
detector responsivity derived from scatter of the response of the individual
detector pixels to the background.

\begin{table}[htb]
\caption{The photometry of NGC~891, as derived from the mosaic maps}
\begin{tabular}{c|ccccc}
\hline\hline
filter & $F$ & $\epsilon(F)$ & $B_{\rm peak}$ & ${\epsilon}(B_{\rm peak})$ 
& ${\epsilon}_{\rm syst}$\\
       & Jy  & Jy   & MJy/sr   & MJy/sr & $\%$                   \\
\hline
C160 & 193 & 15 & 238.7 & 0.7 &  5\\
C200 & 165 &  6 & 195.1 & 1.3 & 10\\
\hline
\end{tabular}
\end{table} 

\section{Comparison between the data and the model for the FIR emission 
of NGC~891}

Especially because it is so well observed, but also because of its edge-on
geometry, which allows the study of the vertical structure of the different
stellar, dust and gas components, NGC~891 was chosen as a prototype since the
study of van der Kruit \& Searle (1981), Bahcall \& Kylafis (1985) and Kylafis
\& Bahcall (1987). More recently, Xilouris et al. (1998, 1999) fitted the 
optical and NIR images of NGC~891 with simulated maps produced by radiative 
transfer calculations, in an attempt to derive the scale lengths and 
heights of the stellar and dust distributions. All these studies concentrated
on modelling the optical appearance of NGC~891.

Popescu et al. (2000a) developed a modelling technique for the whole spectral 
energy distribution, from the UV to FIR and sub-mm, which was tested and also  
first applied to NGC~891. This model successfully reproduced the observed IRAS
and sub-mm flux densities of NGC~891. Furthermore, it is the only model which 
makes direct predictions for the spatial distribution of the FIR
emission. Before comparing the predictions of this model with the new ISOPHOT
data, we review the main characteristics of the dust and stellar distributions
in NGC~891, as derived from this model.

Both a diffuse component and a clumpy component associated
with the star forming regions are calculated. The model for the diffuse component
includes a consistent treatment of grain heating and emission, solves 
the radiation transfer problem for a finite disk and bulge, and 
self-consistently calculates the stochastic heating of grains placed in the
resulting radiation field. 
Compared with the other self-consistent treatments of the 
UV/optical--FIR/submm SEDs our model is unique in that it
analyses the surface brightness distribution
of optical and NIR images to constrain the intrinsic
distributions of the old stellar populations and associated dust (see also
Misiriotis et al. 2001).
The appearance of the optical/NIR images was produced in
an iterative optimisation procedure using the technique for
solving the radiation transfer equation for direct and multiply scattered
light for arbitrary geometries by Kylafis \& Bachall (1987). 
The calculation was done independently for a number of optical and NIR 
images, thus determining the extinction law for diffuse dust empirically.

The ``young'' stellar population was then determined primarily from the
FIR/submm data. The emissivity of the ``young stellar disk'' was
parameterised in terms of the current star formation rate ($SFR$).
A second dust disk of grain mass $M_{\rm dust}$ was associated with the 
young stellar population.
For the clumpy component a third primary parameter, $F$, was introduced to 
denote the fraction of non-ionising UV
which is locally absorbed in HII regions around the massive stars.
These three parameters -$SFR$, $M_{\rm dust}$ and $F$ - are sufficient to fully
determine the FIR-submm SED, due to the precise constraints on the distribution of
stellar emissivity in the optical-NIR and on the distribution and opacity
of dust in the ``old dust disk'' yielded by the radiation 
transfer analysis of the highly resolved optical-NIR images, coupled with the
simple treatment of the young stellar population and associated dust.

Comparison with IRAS and submm data yielded a best solution for NGC~891 with
$SFR=3.8\,M_{\odot}$/yr, $F=0.22$ and $M_{\rm dust}=7\times
10^7\,M_{\odot}$. This corresponds to a disk
central face-on opacity in the V band of ${\tau}^f_V=3.1$ and a
non-ionising UV luminosity of $\sim 8.2\times 10^{36}$\,W. The
luminosity of the diffuse dust emission is $4.07\times10^{36}$\,W, which accounts
for 69$\%$ of the total FIR luminosity, and the luminosity of dust in the
clumpy component is $1.82\times 10^{36}$\,W, making up the remaining 31$\%$ of
the FIR luminosity. 
  
\begin{figure}
\includegraphics[scale=0.7]{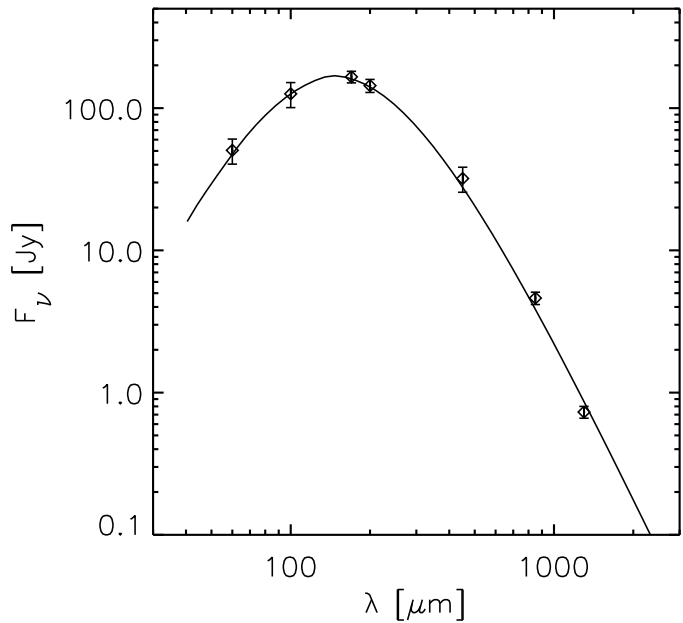}
\caption{
The FIR SED of NGC~891, integrated over $\pm 225^{\prime\prime}$. The measured
points are plotted as diamonds while the solid line represents the FIR 
emission predicted by the model of Popescu et al. (2000a). 
The data points at 170 and 200\,${\mu}$m (derived from our
ISOPHOT maps by integrating between $\pm 225^{\prime\prime}$) constitute 86 
and 87$\%$ of the total flux densities, respectively. The data points at 60, 
100, 450 and 850\,${\mu}$m are taken from Alton et al. (1998) and the data 
point at 1300\,${\mu}$m is taken from Gu\'elin et al. (1993).}
\end{figure}

\subsection{The integrated FIR flux densities}

The predicted FIR SED of NGC~891 is presented in Fig.~2, together with the
IRAS, SCUBA (Alton et al. 1998), IRAM (Gu\'elin et al. 1993) and our ISOPHOT
data. Both the model and the data were integrated
only within $\pm 225^{\prime\prime}$, in order to match the SCUBA and the IRAS 
flux densities derived by Alton et al. (1998). At 170 and 200\,${\mu}$m the 
integrated flux densities within $\pm 225^{\prime\prime}$ are 166 and
144\,Jy, respectively. Our model SED, derived and 
checked on IRAS and submm data by Popescu et al. (2000a), is found to 
be in excellent agreement with the ISOPHOT flux densities as well.

\subsection{The FIR maps}

A more stringent test of the model is to compare its predictions for the
morphology of the dust emission with the observed morphology. 
For this purpose simulated FIR maps were produced using the actual pointing 
data to scan the diffuse disk model. The model map was then convolved with 
empirical PSFs 
derived from point source measurements. The comparison between the observed 
maps at 170 and 200\,${\mu}$m (Fig.~3a and 3b, respectively) and the 
simulated maps at the same wavelengths (Fig.~3c and 3d, respectively) show a 
remarkable agreement.
To search for small differences between the model and the observations, not 
detectable in the maps due to the high dynamical range of the displayed data, 
we present in Fig.~3e and 3f the residual maps of the difference between the 
observed and the simulated maps at 170 and 200\,${\mu}$m, respectively. 

At 170\,${\mu}$m, the main feature in the 
residuals is a localised, unresolved source in the northern side of the disk, 
with a peak of 52.3 MJy/sr. This localised source is probably  
a giant molecular cloud complex - associated with one of the spiral arms,
and not considered in the simulated map, which only includes the diffuse
component of the model. At this FIR wavelength our model predicts an $11\%$
contribution from the star-forming complexes. The integration of the unresolved
source gives a flux density of 13.6\,Jy, which is $7\%$ of the total 
flux density. Furthermore a faint 
extended source  is seen in the southern
side of the galaxy, of 9.5\,Jy integrated flux density. This makes another 
$5\%$ of the total integrated emission. Thus the faint localised sources seen 
in the residual maps sum up to $12\%$ of the total FIR emission. This is 
in reasonable agreement with the prediction of our 
model, which reassures us that the template used in our model and scaled 
according to our model parameters ($SFR$ and the $F$ factor), is indeed a 
good representation for the galaxy. Apart from the two 
sources, a faint extended halo (at $\sim1\%$ brightness level) is seen 
extending at large heights perpendicular to the disk. 

 At 200\,${\mu}$m, the residual map (Fig.~3f) shows the same localised sources
present at 170\,${\mu}$. This time the northern localised source has a flux
density of 8.4\,Jy (5.1$\%$ of the total emission) and the southern localised
source has a flux density of 6.9\,Jy (4.2$\%$). Overall the percentage
contribution of both localised sources to the total flux density is smaller at
200\,${\mu}$m than it is at 170\,${\mu}$m, which means that the localised
sources are warmer than the diffuse component. This is to be expected if we
consider that the localised sources are associated with the HII regions, and
thus consistent with the model of Popescu et al. (2000a).

\begin{figure*}[htp]
\includegraphics[scale=0.7]{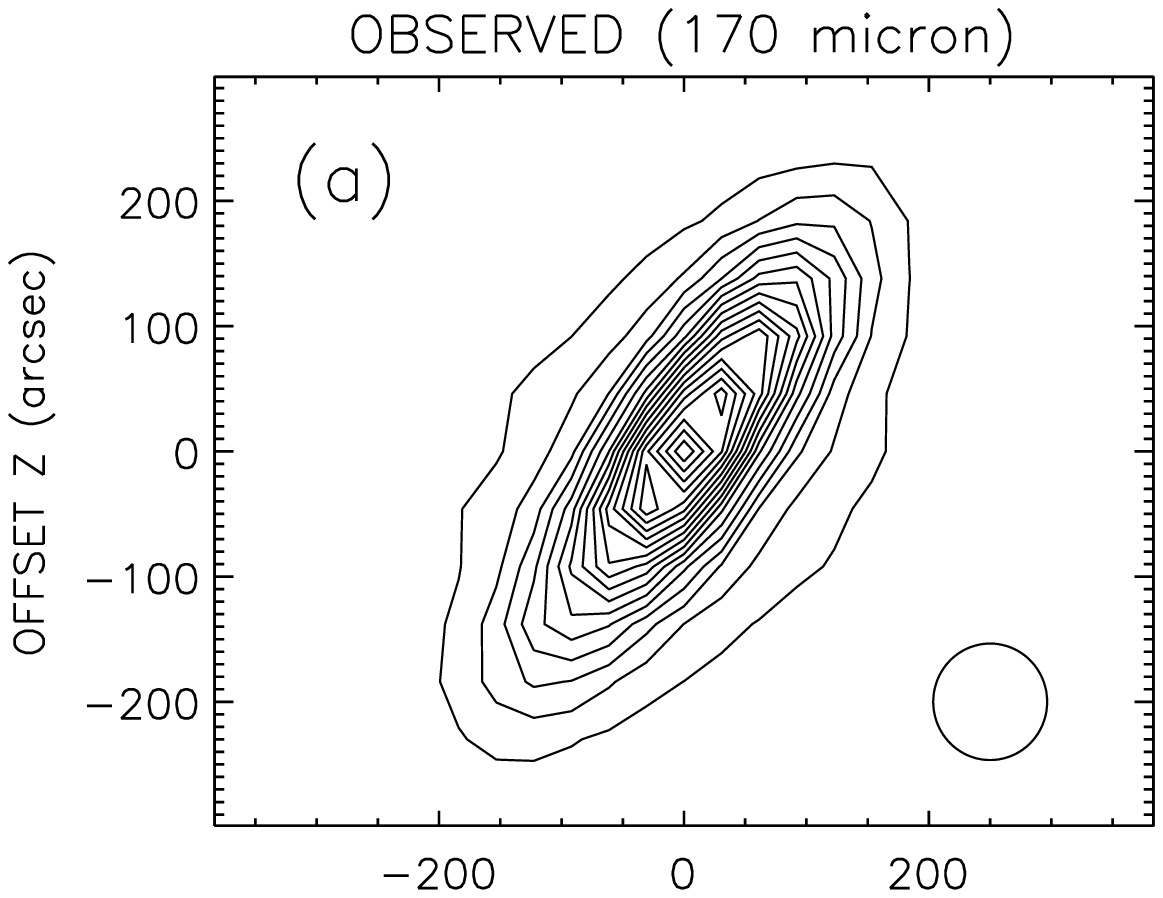}
\includegraphics[scale=0.7]{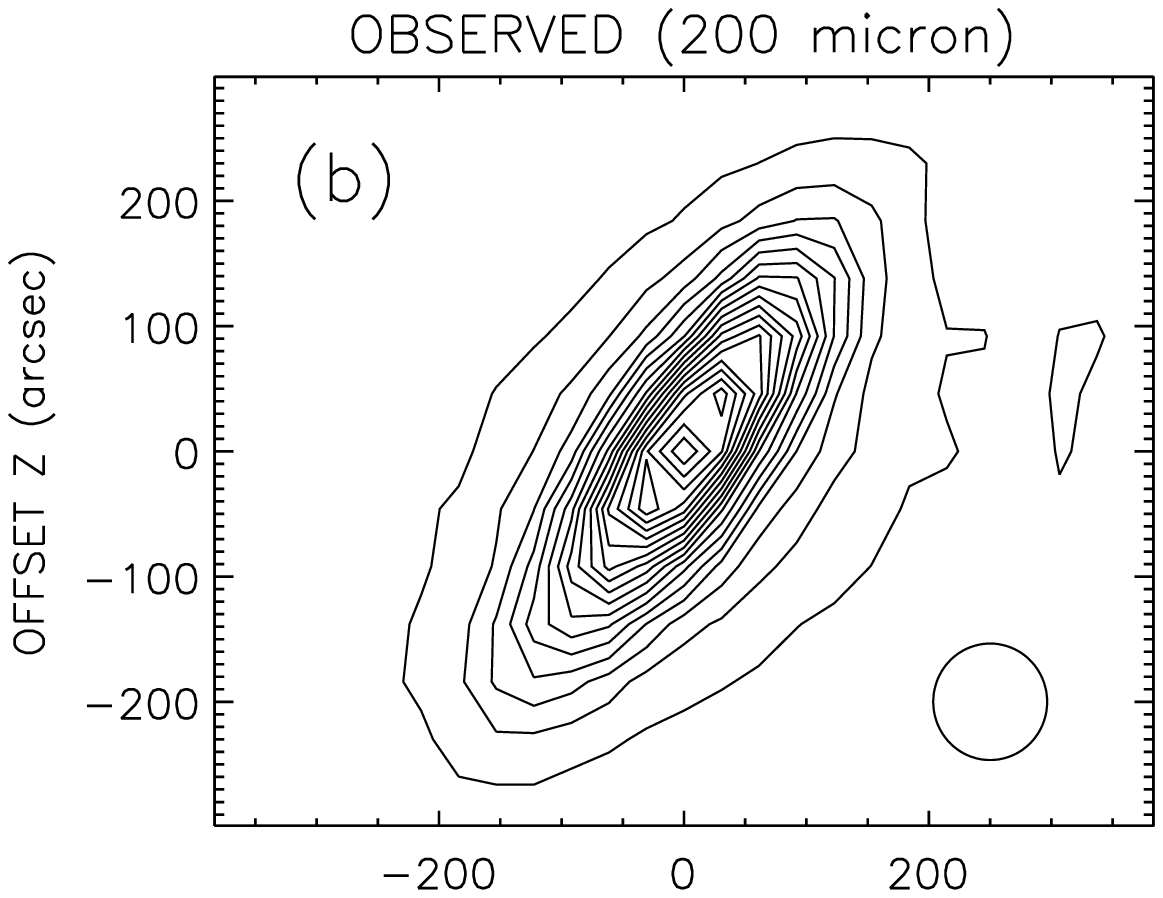}

\vspace*{0.5cm}

\includegraphics[scale=0.7]{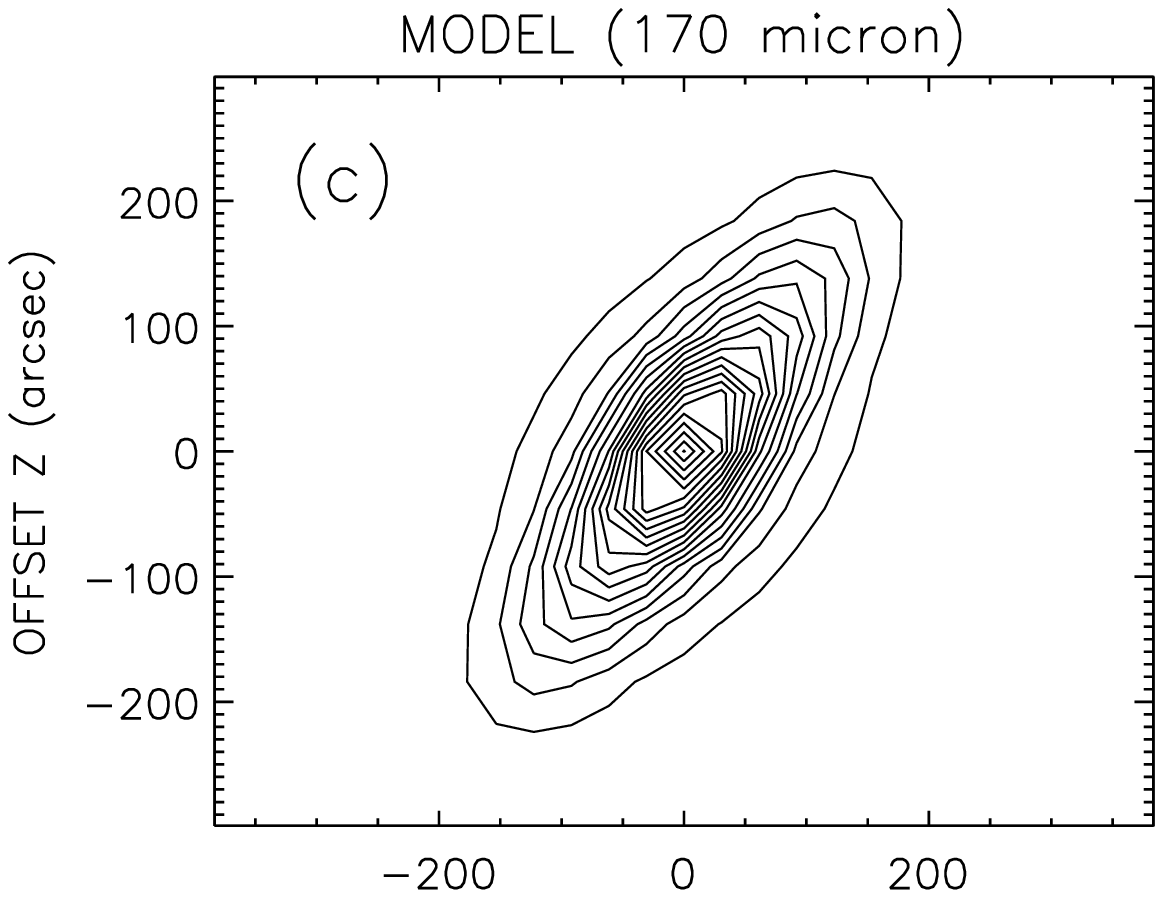}
\includegraphics[scale=0.7]{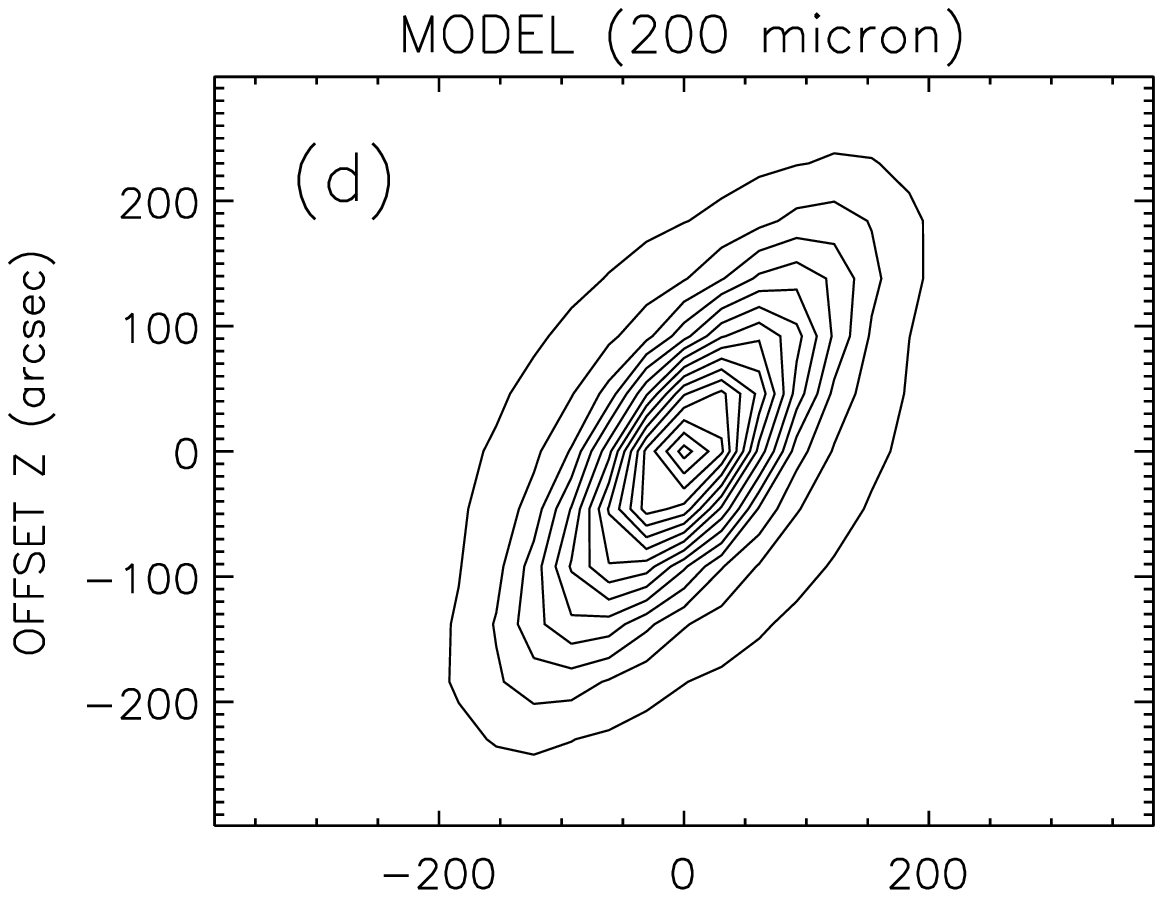}

\vspace*{0.5cm}

\includegraphics[scale=0.7]{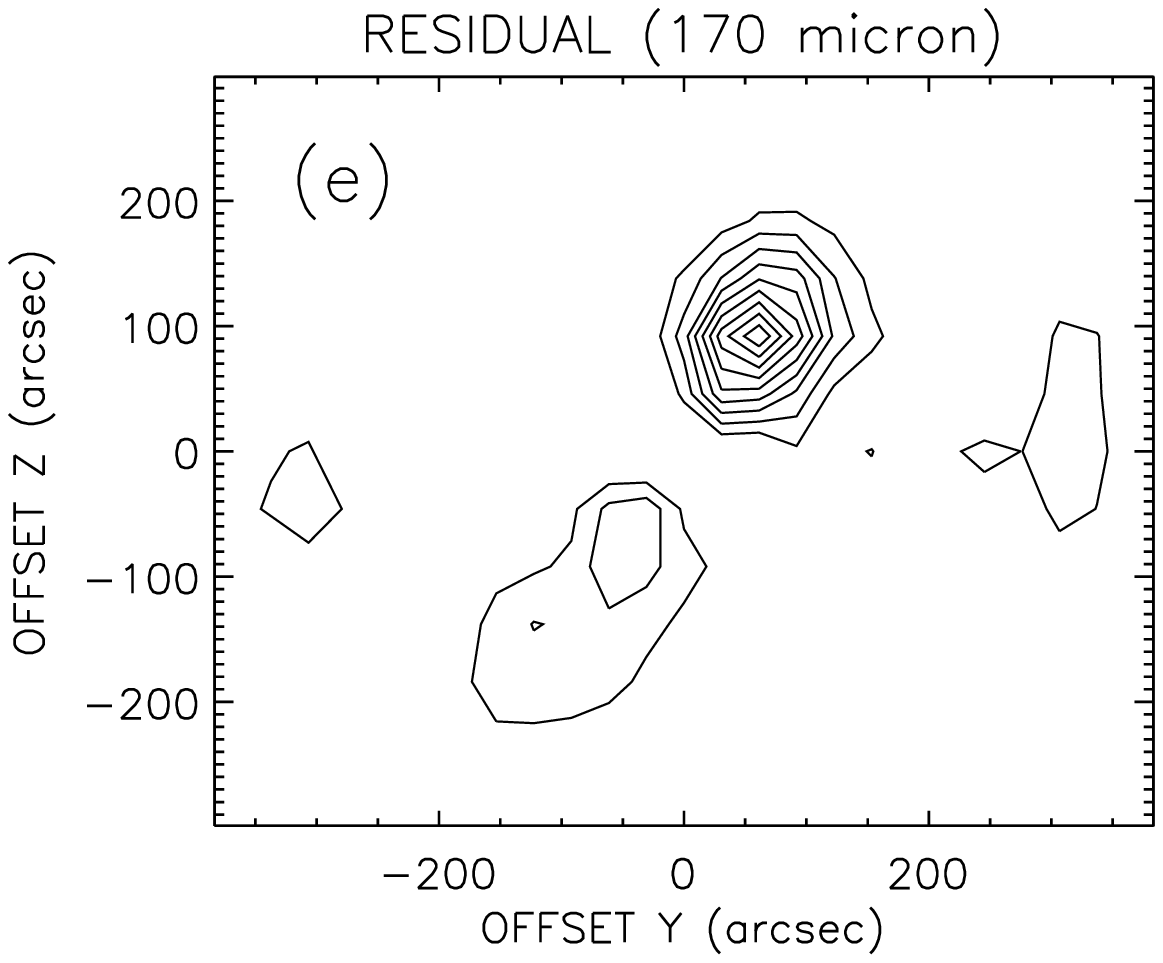}
\includegraphics[scale=0.7]{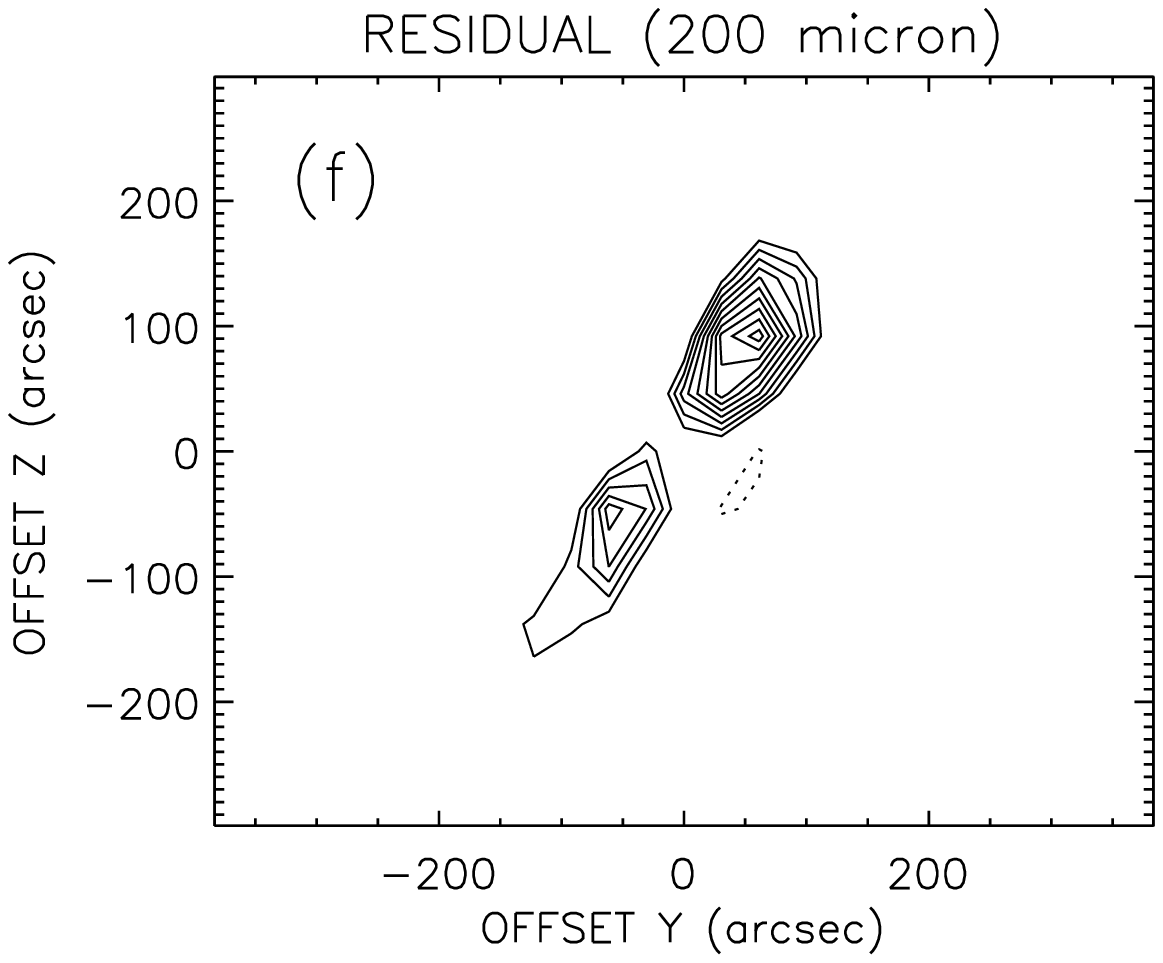}
\caption{ 
a) Contour plot of the observed brightness distribution at
170\,${\mu}$m (in spacecraft coordinates Y and Z). The contours are plotted 
from 11.7 to 226.1\,MJy/sr in steps 
of 12.6\,MJy/sr. The circle shows the perimeter (to the FWHM) of the 
ISOPHOT beam at 170\,${\mu}$m.
b) Contour plot of the observed brightness distribution at
200\,${\mu}$m. The contours are plotted from 7.1 to 188.0\,MJy/sr in steps of 
10.4\,MJy/sr. The circle shows the perimeter (to the FWHM) of the ISOPHOT 
beam at 200\,${\mu}$. 
c) Contour plot of the simulated diffuse brightness distribution at
170\,${\mu}$m. The contour levels are as in panel a).
d) Contour plot of the simulated brightness distribution at
200\,${\mu}$m. The contour levels are as in panel b).
e) Contour plot of the observed minus simulated diffuse brightness 
distribution at 170\,${\mu}$m. The contours are plotted from 
7.5 to 47.5\,MJy/sr in steps of 5.0\,MJy/sr.
f) Contour plot of the observed minus simulated diffuse brightness 
distribution at 200\,${\mu}$m. The contours are plotted from 10.5 to 
37.5\,MJy/sr in steps of 3.0\,MJy/sr. The negative contour (dotted line) 
is at -10.5 MJy/sr.}
\end{figure*}

\subsection{The FIR averaged profiles}

\begin{figure*}[htp]
\subfigure[]{
\includegraphics[scale=0.68]{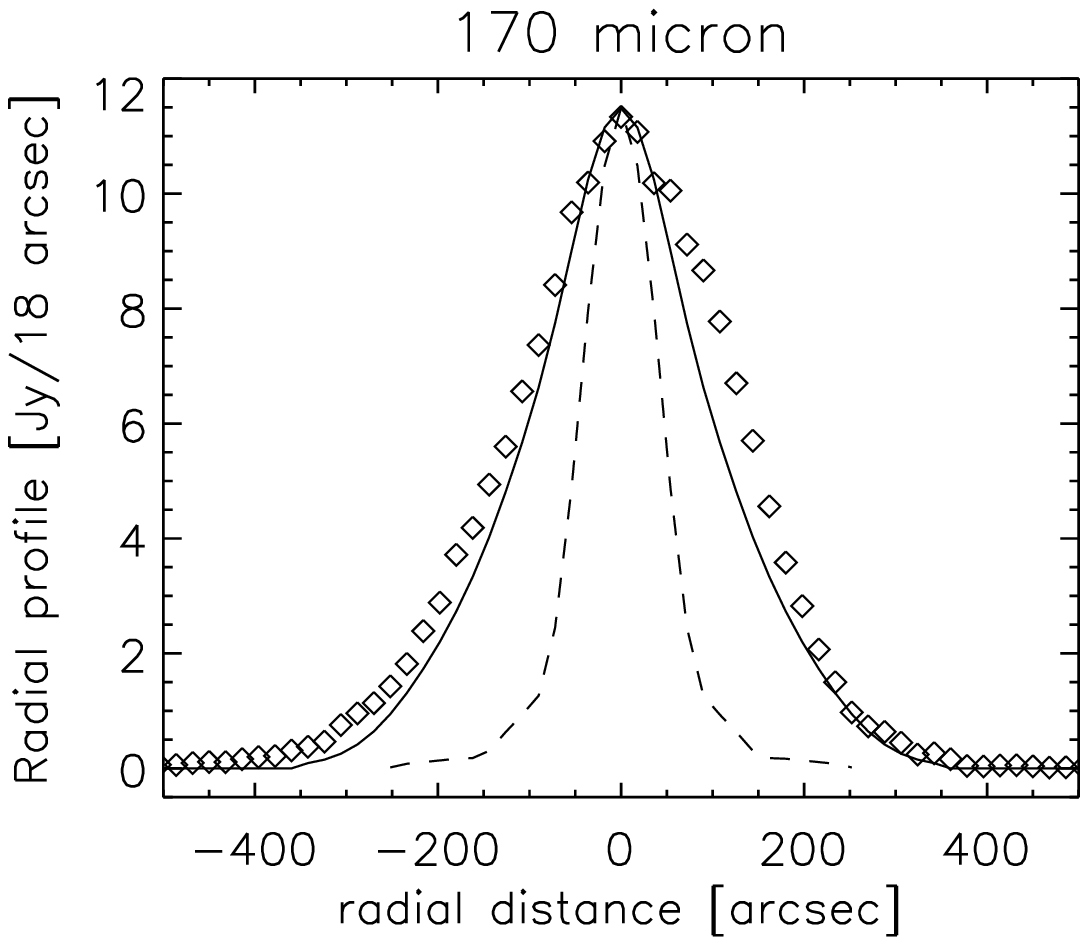}}
\subfigure[]{
\includegraphics[scale=0.68]{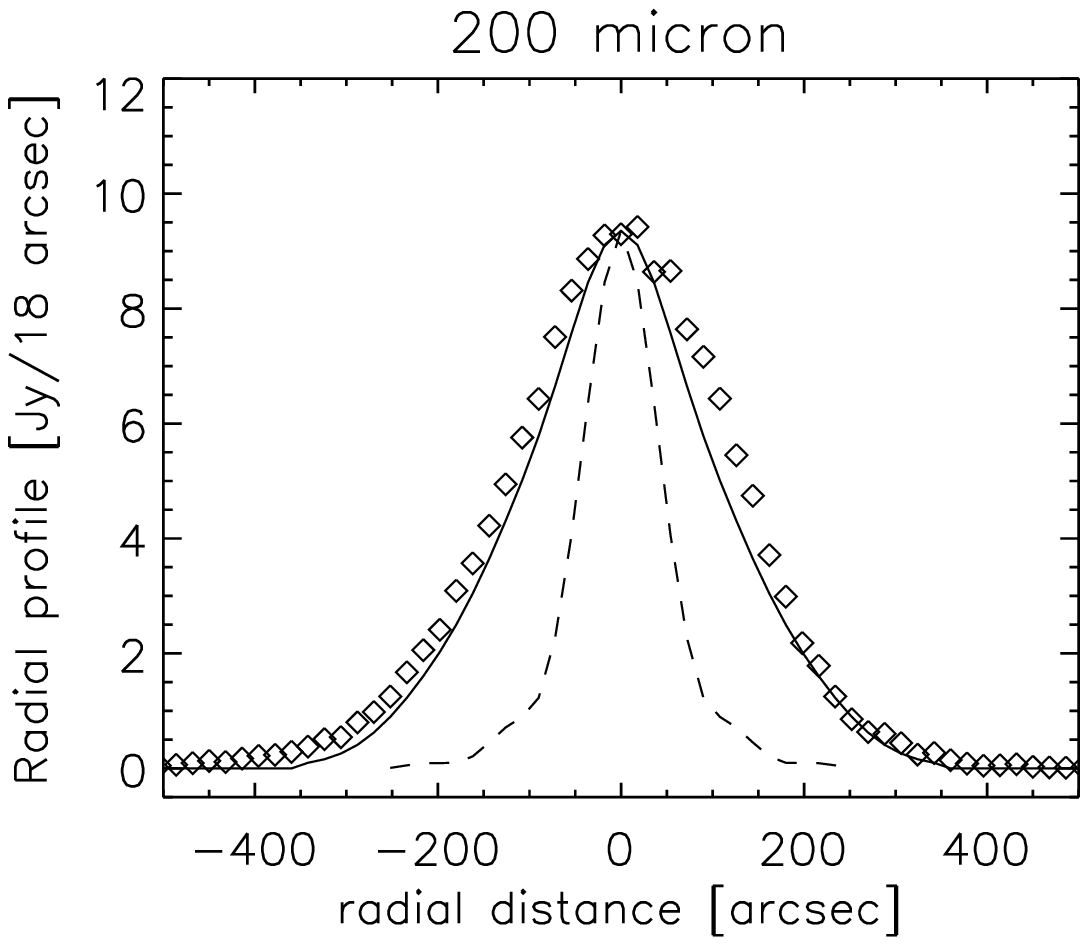}}
\subfigure[]{
\includegraphics[scale=0.68]{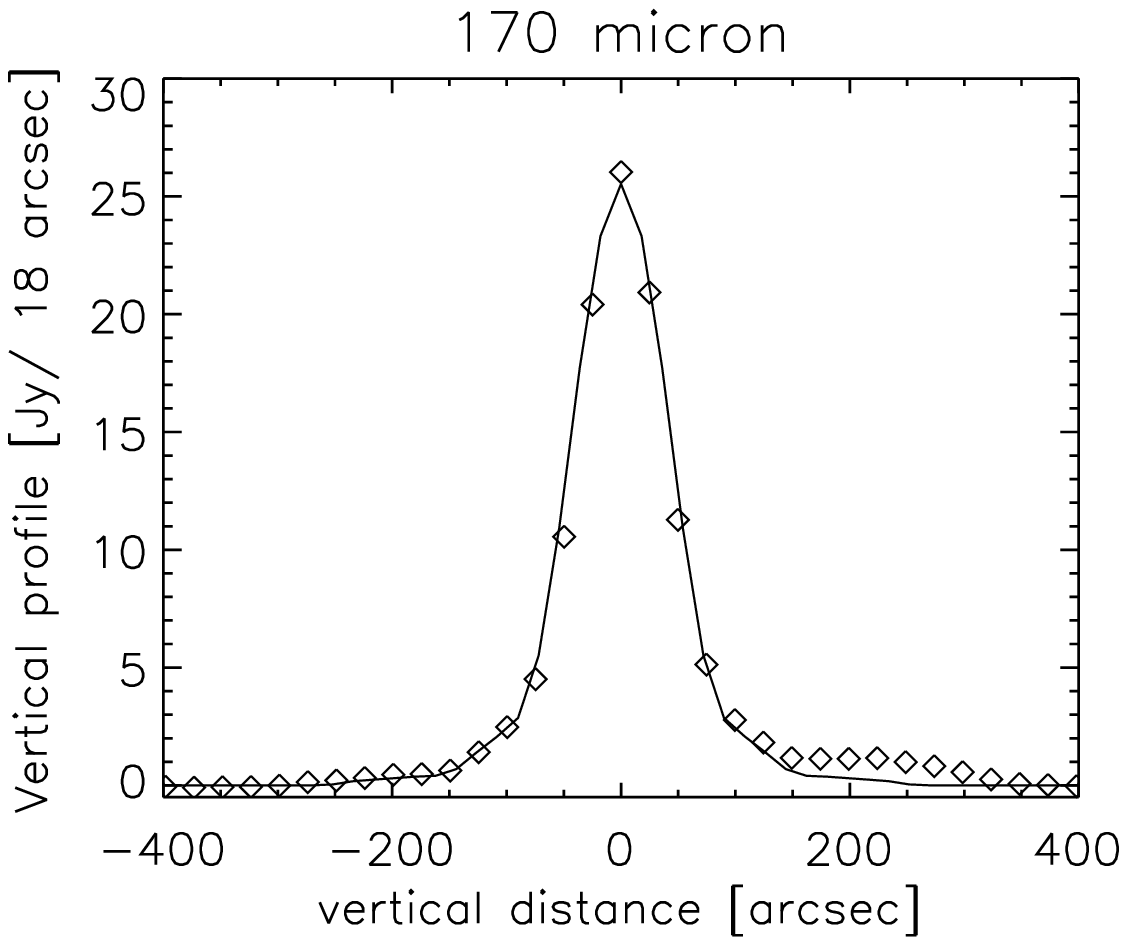}}
\subfigure[]{
\includegraphics[scale=0.68]{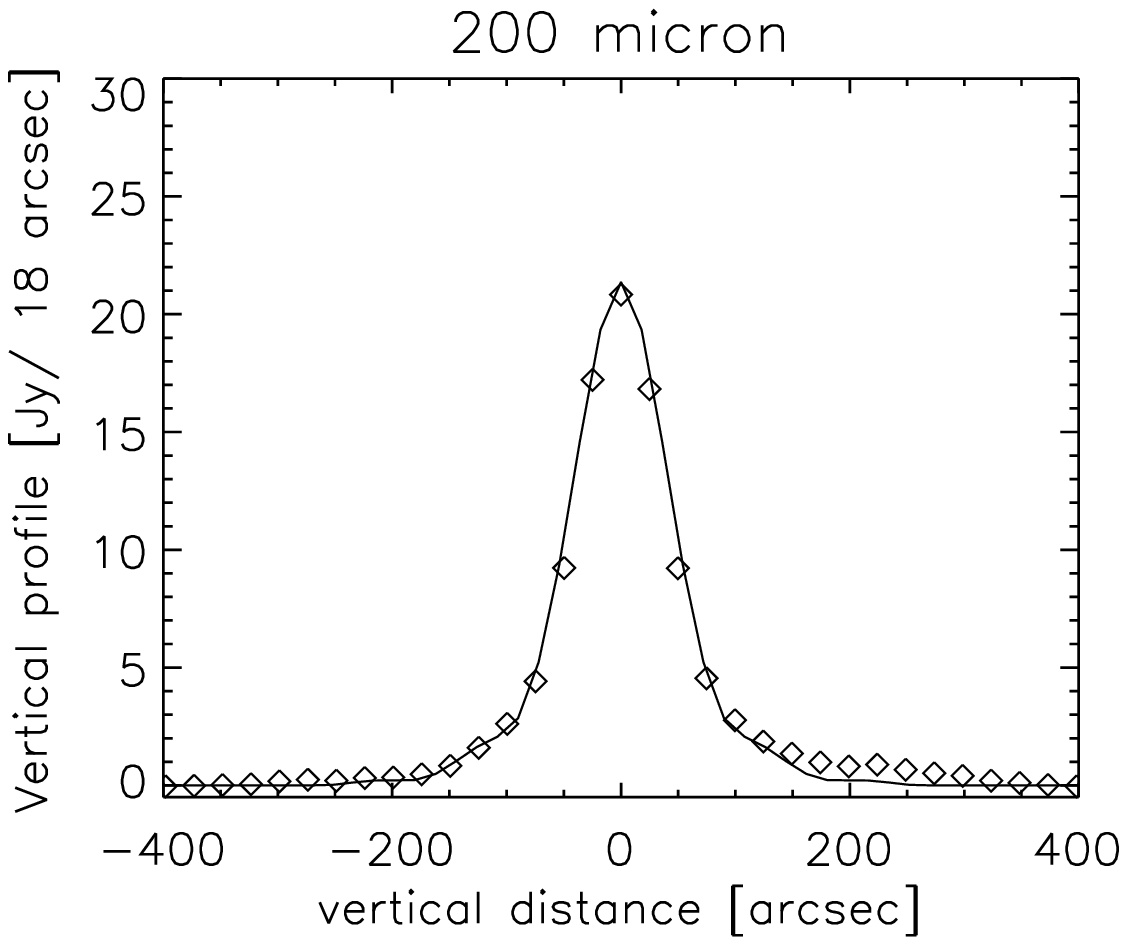}}
\hspace*{5cm}
\subfigure[]{
\includegraphics[scale=0.68]{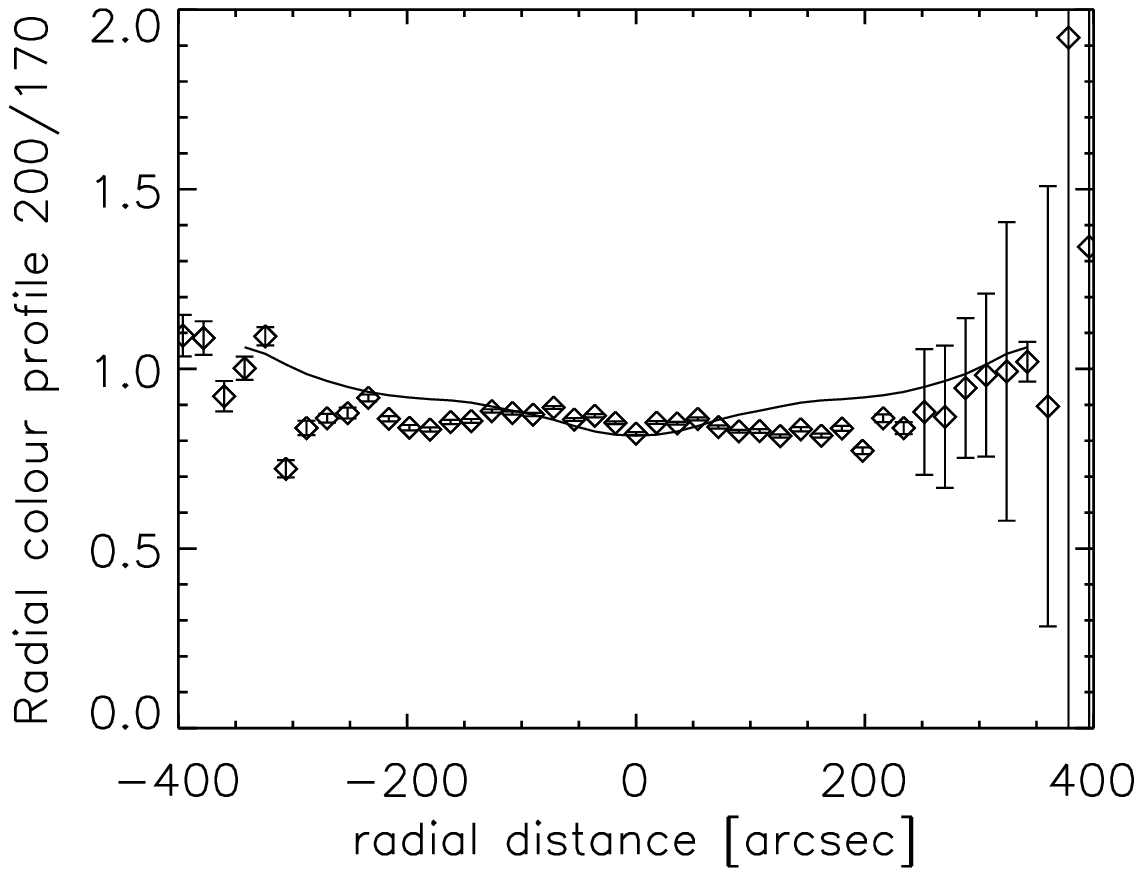}}
\caption{a-b: The radial profiles at 170 and 200\,${\mu}$m produced by 
integrating the emission parallel to the minor axis of the galaxy for each bin 
along the major axis. c-d: The vertical profiles at 170 and 200\,${\mu}$m 
produced by integrating the emission parallel to the major axis of the galaxy 
for each bin along the minor axis. e: The radial colour profile 
F200/F170. The sampling of the profiles is 18 arcsec. 
Solid line: model prediction; diamonds: observed profile; dotted 
line: beam profile.}
\end{figure*}

Another way to compare the predictions of the model with the data is to look at
the averaged profiles, both in the radial and vertical direction. Firstly, 
profiles were produced from the observed maps and from the maps of the 
predicted diffuse emission (Fig.~3a,b,c,d), by integrating the flux parallel 
to the minor axis of the galaxy for each bin along the major axis. The 
resulting profiles, which we refer to as ``radial profiles'', are given 
in Fig.~4a and 4b in the 170 and 200\,${\mu}$m bands, respectively. Again, a 
good agreement between the data points and the 
model prediction is apparent at both wavelengths. Centred at +100 arcsec 
there is an excess emission in the observed profile with respect to the 
predicted one, which is due to the localised source in the northern half of 
the galaxy, already identified in the residual maps (Fig.~3e,f). This asymmetry
is also seen in the 850\,${\mu}$m map of Alton et al. (1998) and in the 
1300\,${\mu}$m map of Gu\'elin et al. (1993). Likewise, the
fainter localised source in the southern half of the galaxy manifests itself
through the excess observed emission seen around -150 arcsec radius. At both
wavelengths there is some faint emission apparent in the observed profile
extending beyond the edge of the optical disk (360 arcsec). This emission 
is considered in Popescu \& Tuffs (2003).

Similar to the construction of the radial profiles, profiles perpendicular
to the major axis of the galaxy 
were calculated by integrating the flux parallel to the major axis of the
galaxy for each bin along the minor axis. The resulting profiles, which we
refer to as ``vertical profiles'', are given in
Fig.~4c,d. As expected, the predicted disk emission is unresolved by ISOPHOT,
and therefore the predicted profile coincides with the beam profile. The
observed profile is also unresolved up to $\pm 100$ arcsec, which corresponds
to a factor of 10 in dynamic range. At larger heights
above the disk there is some faint diffuse emission (at a level of a few
percent of the peak brightness) in excess of the model
predictions for the disk. We have already discussed in Sect.~2 that this
emission may be of an instrumental nature, although a detection of extraplanar
emission cannot be discounted completely. Such an emission may be expected to
arise from dust grains carried out by a gentle wind, as described in Popescu et
al. (2000b).

The measured and predicted radial colour profiles 200/170 are plotted in 
Fig.~4e. The predicted profile has a smooth progression towards colder emission
with increasing radial distance. The measured profile broadly follows this
prediction, although there may be evidence for a tendency for a 
flatter profile within $\pm 200$. The larger error bars on the data points 
beyond +200 arcsec are due to the lower brightness of the 170\,${\mu}$m outer 
disk on the northern side.

\section{Discussion}

At the wavelength of the ISOPHOT measurements presented here the model for
NGC~891 predicts that the bulk of the FIR dust emission is from the diffuse
component. The close agreement between the data and the model predictions, both in 
integrated flux densities, but especially in terms of the spatial distribution,
constitutes a strong evidence that the large scale distribution of stellar
emissivity and dust predicted by the model is in fact a good representation of
NGC~891. In turn, this supports the prediction of the model that the dust
emission in NGC~891 is predominantly powered by UV photons. 

Depending on the FIR/submm  wavelength, the UV powered dust emission arises in 
different
proportions from within the clumpy component and from the diffuse component.  
For example at 60\,${\mu}$m, 61$\%$ of the FIR emission is powered by UV
photons locally 
absorbed in star forming complexes, 19$\%$ by diffuse UV photons in the weak 
radiation fields in the outer disk (where stochastic emission 
predominates), and 20$\%$ by diffuse optical photons in high energy densities
in the inner part of the disk and bulge. At 100\,${\mu}$m there are
approximately equal contributions from the diffuse UV, diffuse optical and
locally absorbed UV photons. At 170, 200\,${\mu}$m and submm wavelengths, most
of the dust emission in NGC~891 is
powered by the diffuse UV photons. So our analysis does not support the
preconception that the weakly heated cold dust (including the dust emitting
near the peak of the SED sampled by the ISOPHOT measurements presented here) 
should be predominantly powered by optical rather than UV photons. The reason is as
follows: 
the coldest grains are those which are in weaker radiation fields, either in
the outer optically thin regions of the disk, or because they are shielded from
radiation by optical depth effects. In the first situation the absorption
probabilities of photons are controlled by the optical properties of the
grains, so the UV photons will dominate the heating. The second situation
arises for dust associated with the young stellar population, where the UV
emissivity far exceeds the optical emissivity.

\section{Summary}

Here we have presented deep FIR maps of the edge-on spiral galaxy NGC~891,
obtained with the ISOPHOT instrument on board ISO. The observations were done
near the peak of the FIR SED, namely at 170 and 200\,${\mu}$m. Both the
integrated flux densities and the surface brightness distributions obtained 
from our ISOPHOT maps were found to be in excellent agreement with the
 predictions of the model for the optical/FIR/submm SED of
NGC~891 of Popescu et al. (2000a). Furthermore, the model distribution of
emitters and absorbers in NGC~891 is able to reproduce both the
observed optical/NIR images and the FIR maps. NGC~891 is the first 
galaxy for which an intrinsic distribution of dust and stars was derived 
under such strong observational constraints.

\begin{acknowledgements}
We would like to thank the anonymous referee for his useful comments and
suggestions. 
\end{acknowledgements}

\end{document}